\documentclass[]{emulateapj}

\usepackage{natbib}

\defcitealias{bf76}{BF76}
\defcitealias{rsm}{R94}

\begin{document}

\title{Deep 1.4 GHz VLA Observations of the Radio Halo and Relic in Abell 2256}

\author{T. E. Clarke}
\affil{Naval Research Laboratory, Code 7213, 4555 Overlook Ave SW, Washington, DC, 20375}
\affil{Interferometrics Inc., 13454 Sunrise Valley Dr., Suite 240, Herndon, VA, 20171}

\and

\author{T. A. En{\ss}lin}
\affil{Max-Planck-Institut f\"ur Astrophysik, Karl-Schwarzschild-Str.1, Postfach 1317, 85741 Garching, Germany}

\begin{abstract}
We present deep VLA observations of the merging galaxy cluster Abell
2256. This cluster is known to possess diffuse steep spectrum radio
relic emission in the peripheral regions. Our new observations provide
the first detailed image of the central diffuse radio halo emission in
this cluster. The radio halo extends over more than 800 kpc in the
cluster core, while the relic emission covers a region of $\sim 1125
\times 520$ kpc. A spectral index map of the radio relic shows a
spectral steepening from the northwest toward the southeast edge of
the emission, with an average spectral index between 1369 MHz and 1703
MHz of $\alpha=-1.2$ across the relic. Polarization maps reveal high
fractional polarization of up to 45\% in the relic region with an
average polarization of 20\% across the relic region. The observed
Faraday rotation measure is consistent with the Galactic estimate and
the dispersion in the rotation measure is small, suggesting that there
is very little contribution to the rotation measure of the relic from
the intracluster medium. We use these Faraday properties of the relic
to argue that it is located on the front side of the cluster.
\end{abstract}


\section{Introduction}

In the hierarchical model of structure formation, objects form from
the gravitational collapse of initial density enhancements and
subsequently grow through accretion and mergers. Numerical simulations
of structure formation show that clusters of galaxies are found to
preferentially form at the intersections of large filamentary
structures \citep{w91,kw93}. Due to the high-density environment in
which they reside, clusters of galaxies are expected to undergo
several merger events during their lifetime. These merger events are
highly energetic ($10^{63}-10^{64}$ ergs) and thus provide a
significant energy input into the intracluster medium (ICM). Large
scale structure simulations \citep{m00} as well as 3D MHD/N-body
simulations \citep{r99a,r99b,dolag99} find that the shocks and
turbulence associated with a major cluster merger event can
significantly amplify the intracluster magnetic field strength and
accelerate relativistic particles which, in the presence of a magnetic
field, emit synchrotron emission.

Radio observations toward a number of galaxy clusters reveal the
presence of large regions of diffuse radio emission which extend over
scales of $>$ 600 kpc in the ICM and have no obvious optical
counterpart \citep[see review by][]{gf02}. This emission appears to
fall in two categories: {\it halos} which are centrally located in the
cluster, relatively regular in shape, and unpolarized, and {\it
relics} which are peripherally located, fairly elongated and
irregular, and often highly polarized \citep{fg96}. In an X-ray flux
limited cluster sample, \citet{gtf} find that 5\% of clusters contain
a known radio halo and 6\% contain a detected radio relic.

The presence of these large regions of diffuse synchrotron emission
reveals the large scale distribution of relativistic particles and
magnetic fields in the intracluster medium. Despite extensive searches
through the 1.4 GHz NRAO VLA Sky Survey \citep{gtf} and the 330 MHz
Westerbork Northern Sky Survey \citep{ks} as well as a number of
targeted searches for radio halos and relics
\citep{kcecn,gf00,liang00,vbm,rhlp,sr,fbgn,hswd,and88,c85,hsj} there are still
a relatively small number of clusters known which host this diffuse
emission.

Galaxy clusters which are confirmed to contain diffuse radio emission
also show significant evidence of merger activity. The clusters which
contain radio halos tend to be very X-ray luminous, massive clusters
\citep{cola,liang00,buote} which display a significant amount of X-ray
substructure. Although this suggests that the merger event is the
trigger for the diffuse emission, it should be noted that while only
$\sim$ 5\% of all galaxy clusters in a complete X-ray flux limited
sample appear to contain diffuse emission, more than 40\% of clusters
show evidence of merger activity \citep{jf}. On the other hand,
\citeauthor{gf02} noted that the detection rate of diffuse emission is
significantly enhanced for high X-ray luminosity clusters where it
reaches $\sim$33\% for a sample with X-ray luminosity larger than
$10^{45}$ ergs s$^{-1}$. \citet{buote} suggests that this is due to
the preferential formation of diffuse radio emission in massive
clusters which are currently experiencing violent mergers.

In this paper we present new radio maps of the diffuse emission in
Abell 2256. The aim of our investigation is to use this cosmic
laboratory to study details of plasma physics, magneto-hydrodynamics,
and eventually particle physics. In particular, the goal of the
polarimetry observations is to use the Mpc scale radio relic emission
to investigate statistical properties of the ICM magnetic field. This
first paper presents the data and a discussion of the geometry of the
system. The application of statistical tools to the data to extract
information on magneto-hydrodynamical turbulence, particle
acceleration, etc.\ is left to follow-up publications. Unless
otherwise noted, we adopt WMAP cosmological parameters \cite{wmap}
H$_0$=71 km s$^{-1}$ Mpc$^{-1}$, $\Omega_{\Lambda}$=0.73, and
$\Omega_m$=0.27. At the redshift of Abell 2256 (z=0.0594), this
corresponds to a linear scale of 1.13 kpc/arcsec.

\section{Target Cluster: Abell 2256}

Abell 2256 is a rich, nearby ($z=0.0594$) galaxy cluster which was one
of the first targets observed by ROSAT. Analysis of these observations
\citep{b91a} revealed significant X-ray substructure in the
cluster. The X-ray surface brightness distribution shows two separate
X-ray peaks that suggest it is undergoing a merger event. The X-ray
temperature map of the cluster \citep{b94} indicated that the
infalling component is cooler than the main cluster body and that
there are two hot regions roughly perpendicular to the merger
axis. These hot regions appear similar to those seen in simulations of
merger events where the merger has not yet proceeded past core passage
\citep{sm,rlb}. More recent $Chandra$ observations of the cluster by
\citet{sun} provide much more detail on the cluster merger state. They
find a sharp surface brightness feature south of the merging
subcluster. The temperature map of the region shows a temperature jump
across the edge with the cooler gas associated with the
subcluster. \citeauthor{sun} suggest that the structure south of the
subcluster core is similar to the ``cold fronts'' seen in merging
clusters such as Abell 2142 \citep{mm00} and Abell 3667
\citep{vmm01}. These features are thought to delineate the contact
surface between the undisturbed cool core of a merging system and the
hot thermal gas through which they are moving. The $Chandra$ data also
reveal a third component to the X-ray structure in the
system. \citet{sun} find an X-ray ``shoulder'' located 2\arcmin\ east
of the main X-ray peak in the core of the cluster. This structure may
be the remnant of an old merging component in the cluster, or may
represent substructure in the core of the primary.

Radial velocity studies of 277 cluster member galaxies by
\citet{blc02} find that the cluster is composed of three
substructures. The two main galaxy groups correspond spatially to the
primary cluster and the infalling subcluster. The third group is
located to the north of the cluster core and apparently represents a
previously unknown merging system. \citeauthor{blc02} suggest that
the subcluster is infalling onto the primary cluster from the northwest,
while the newly identified group is located on the near side of the
system and is merging from the north. They interpret the radio relic
emission as the result of the merger of the group with the other
systems. A similar distribution of three merging components was also
identified by \citet{miller03}. They suggest a slightly different
scenario where the merging group has already passed through the system
from the south close to the line of sight and is currently located on
the far side of the system. On the other hand, \citeauthor{miller03} point
out that the low velocity dispersion of the group does not appear to
be consistent with a post-merger scenario, although they suggest that
the low velocity dispersion may be an artifact of the galaxy
assignment algorithm.

Radio observations of Abell 2256 \citep{bf76,mm78,bf79,rsm} show that
the cluster is host to a remarkable assortment of sources. There are
at least four head-tail galaxies in the cluster, one of which has a
narrow straight tail that extends for at least 480 kpc
\citep[][hereafter R94]{rsm}. The northwest region of the cluster
contains two large, sharp-edged radio relics which appear to be
roughly at the location of the infalling cooler
subcluster. \citet[][hereafter BF76]{bf76} also suggest the
possibility of a cluster-center radio halo based on Westerbork 610 MHz
observations. These observations are discussed in more detail in
\S~\ref{sect:I_prop}.

\section{Radio Observations and Data Reductions}

Abell 2256 was observed at four frequencies across the 1.4 GHz band
using the National Radio Astronomy Observatory's VLA\footnote{NRAO is
operated by Associated Universities Inc., under a cooperative
agreement with the National Science Foundation.} in the C and D
configurations. The observational parameters are summarized in
Table~\ref{tbl:data}. All observations include full polarization
information and used 3C286 as the primary flux density calibrator and
absolute electric vector position angle calibrator. Phase calibration
was obtained from the nearby calibrator 1803+784, and multiple
observations of 1458+718 over a large range of parallactic angles were
used to calibrate the polarization leakage for each
antenna. Observations at 1703 MHz on 1999 April 29 were severely
impacted by radio frequency interference during the first 1.5
hours of observation and this data has been removed from the analysis.

Data were calibrated and reduced using NRAO's Astronomical Image
Processing System (AIPS). The data were calibrated following standard
procedures. Wide-field, 3-D imaging techniques were implemented
within the AIPS task IMAGR to allow us to compensate for the
non-coplanar geometry of the VLA \citep{cp92}. Imaging used a total 22
facets to cover the primary beam and bright outlying sources. We also
included the multiresolution clean option within the AIPS IMAGR task
to allow us to better image the complex diffuse structure in this
system. Each data set was self-calibrated using phase-only followed by
amplitude and phase solutions. The final images used for spectral
index and Faraday rotation measures were created using a suitable
taper within the AIPS task IMAGR followed by a convolution with a
Gaussian to exactly match each beam. The polarized intensity,
fractional polarization, and polarization position angle images were
created from the stokes Q and U images at a given frequency. The
Faraday rotation measure (RM) image was determined from the four
frequencies around 1.4 GHz and provides an unambiguous measurement for
${\rm |RM|}< 300$ rad m$^{-2}$

\section{Total Intensity Properties}
\label{sect:I_prop}

Previous radio observations of Abell 2256 \citep{bf76,rsm,miller03}
have revealed that it contains very complex radio
emission. \citetalias{bf76} attached alphabetic designations to numerous
radio sources in the cluster. These designations were expanded in
subsequent papers by \citet{bf79}, and \citetalias{rsm} as more sensitive
and higher resolution data became available. \citetalias{rsm} designate 34
individual sources in the direction of the cluster. We will not
discuss the majority of the sources in this paper as we concentrate on
lower resolution, high surface brightness sensitivity images of the
extended diffuse emission. A detailed discussion, including high
resolution radio overlays on optical images of the sources is given
in \citet{miller03}. Combining their optical data with that of
\citet{blc02}, \citeauthor{miller03} have velocity measurements for 49
of the 54 candidate cluster radio galaxies and find that 40 of these
are confirmed cluster members.

\begin{figure}
\vskip 2.4truein
\includegraphics{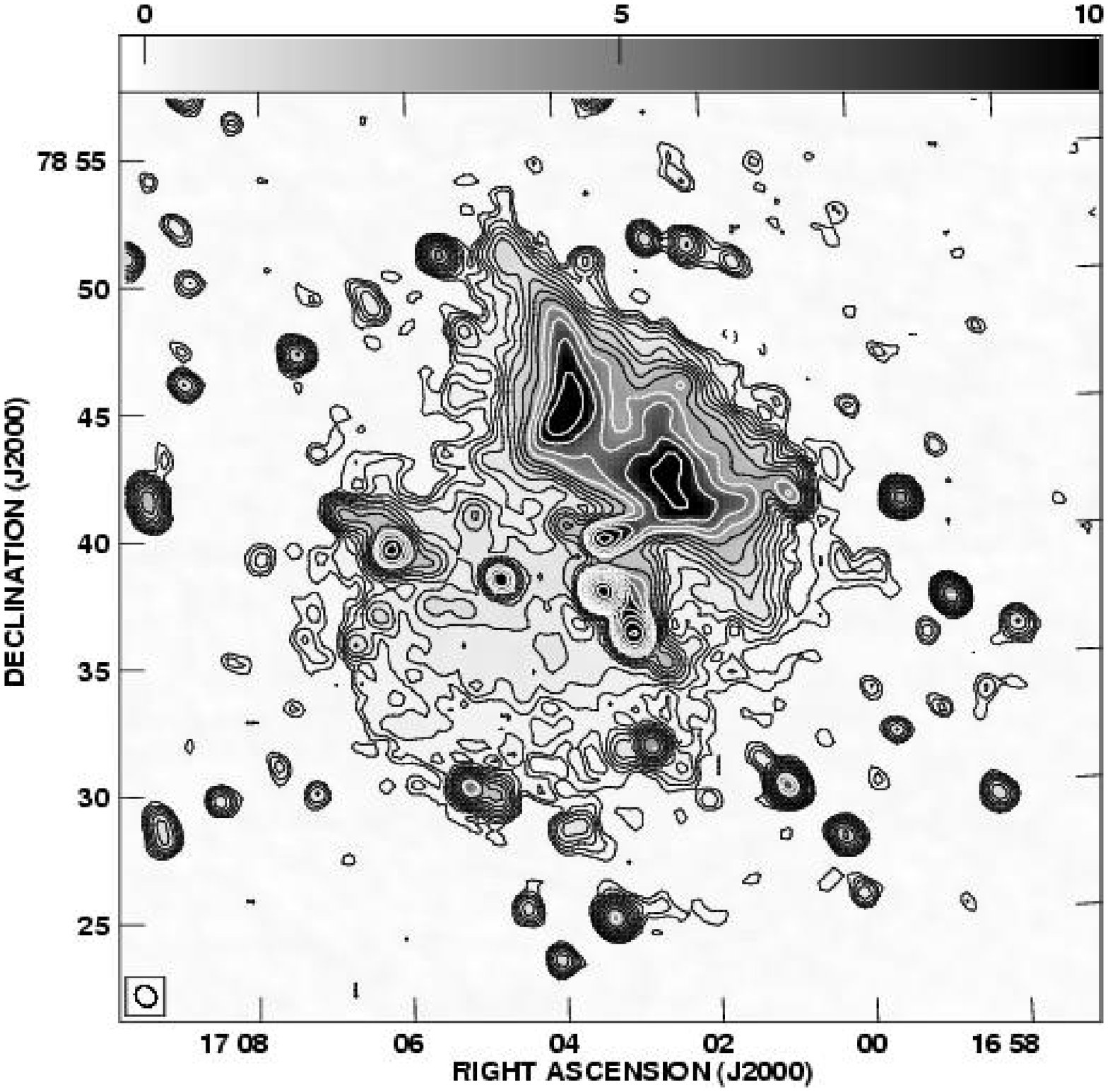}
\includegraphics{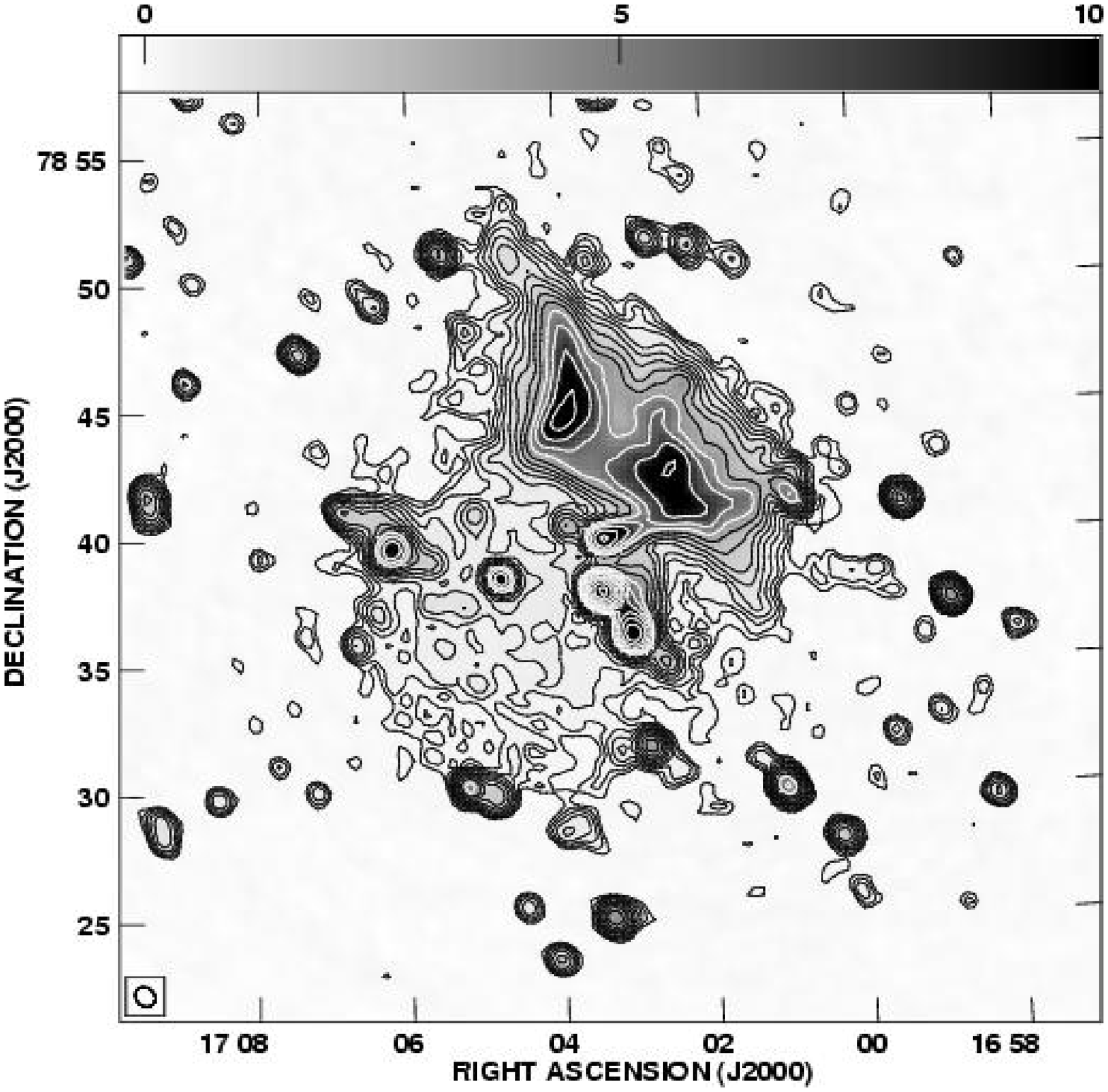}
\includegraphics{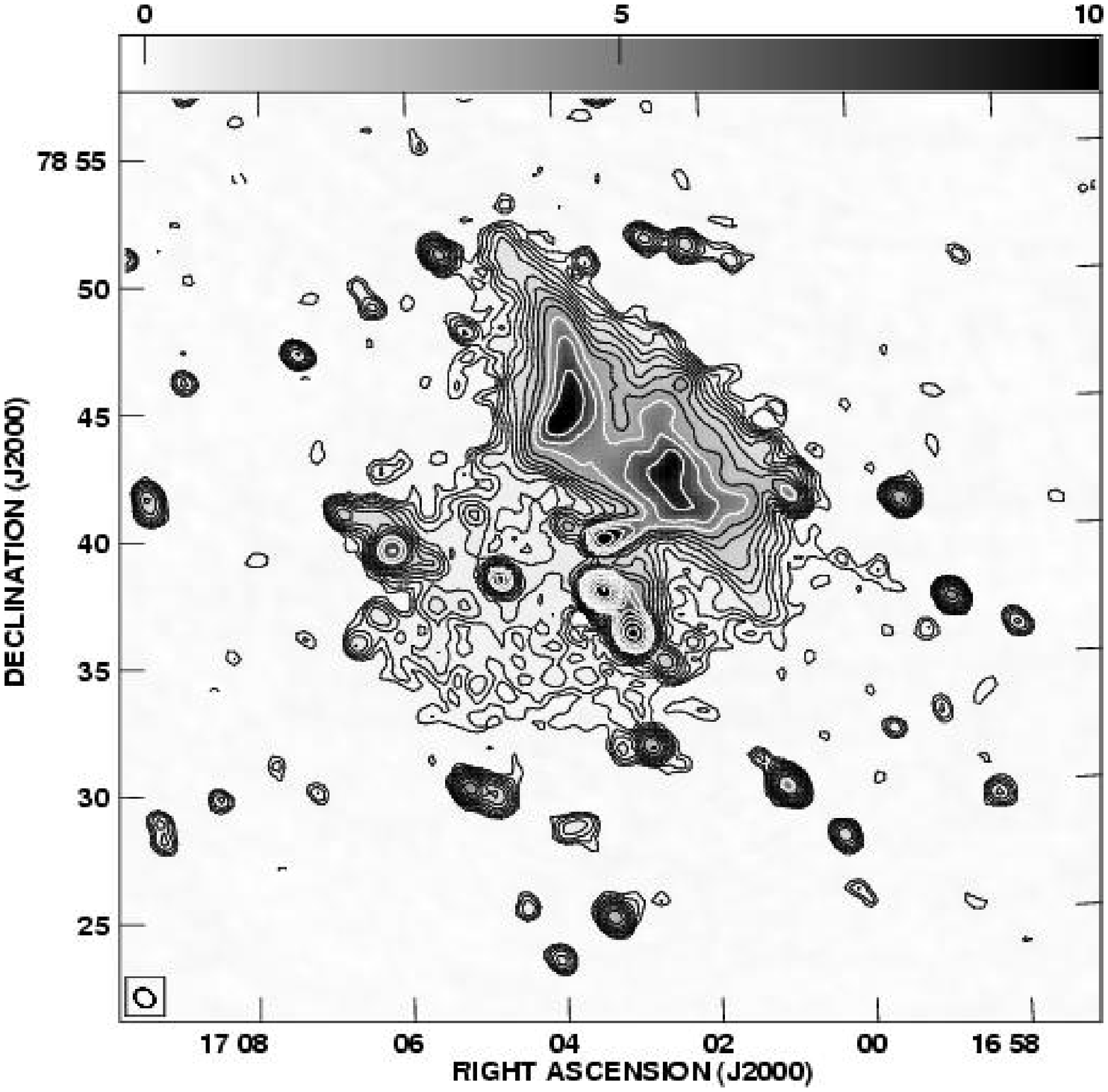}
\includegraphics{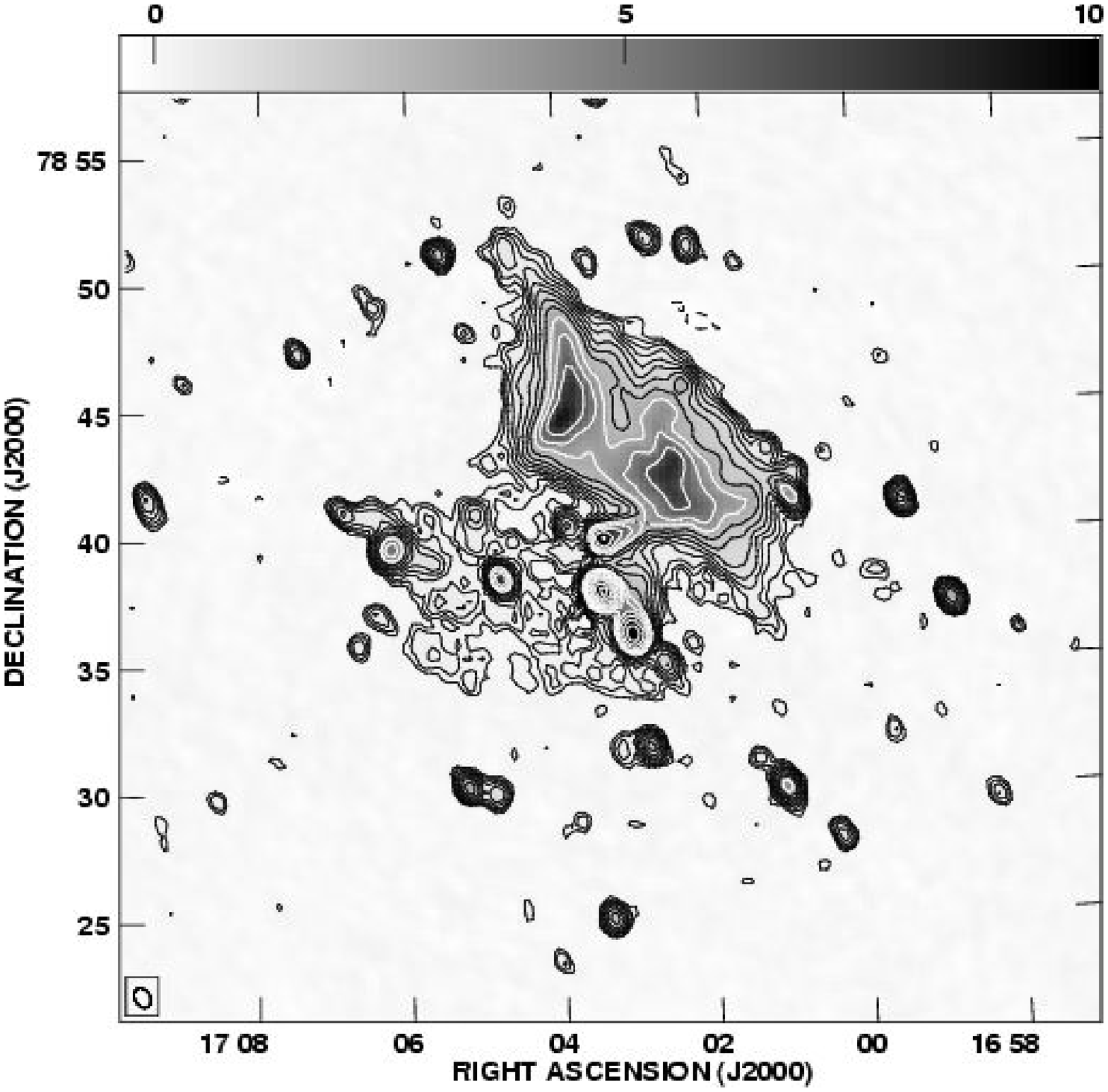}
\vskip 5.7truein
\caption{VLA D configuration total intensity greyscale and contours of
Abell 2256. In addition to the compact sources, the diffuse emission
associated with the peripheral relics and central radio halo is
visible. All contours are plotted as -3, 3, ... times the rms level
at $\sqrt{2}$ intervals. a) 1369 MHz contours. The rms level of 59
$\mu$Jy beam$^{-1}$ and the restoring beam is $\sim$
52\arcsec$\times$45\arcsec. b) 1417 MHz contours with an rms level of 62
$\mu$Jy beam$^{-1}$ where beam is $\sim$ 50\arcsec$\times$43\arcsec. c)
1512.5 MHz contours with an rms level of 60 $\mu$Jy beam$^{-1}$ and a
beam of $\sim$ 49\arcsec$\times$39\arcsec. d) 1703 MHz contours with an rms
level of 68 $\mu$Jy beam$^{-1}$ and a beam of $\sim$
48\arcsec$\times$36\arcsec. 
}
\label{fig:vla_l}
\end{figure}

In the top left panel of Figure~\ref{fig:vla_l} we show the 1369 MHz D
configuration image of Abell 2256 covering a region of $2.6 \times
2.5$ Mpc. The well-known radio relic region \citepalias[G and H
in][]{bf76} is visible as the bright elongated region to the
north-west. Our observations show the first clear detection of
additional diffuse emission (at a much lower surface brightness)
centered to the south-east of the relics. This emission is co-incident
with the diffuse emission seen around source D by \citetalias{bf76}
and is referred to as the halo below. A third region of interest in
Abell 2256 is associated with the Z-shaped source F in the
\citetalias{bf76} notation. This source is visible on the
north-eastern edge of the radio halo and has been noted for its
unusual shape and spectral properties in several papers
\citep{mm78,bf79,rsm,miller03}.

\subsection{Relic Emission}

\begin{figure}
\plotone{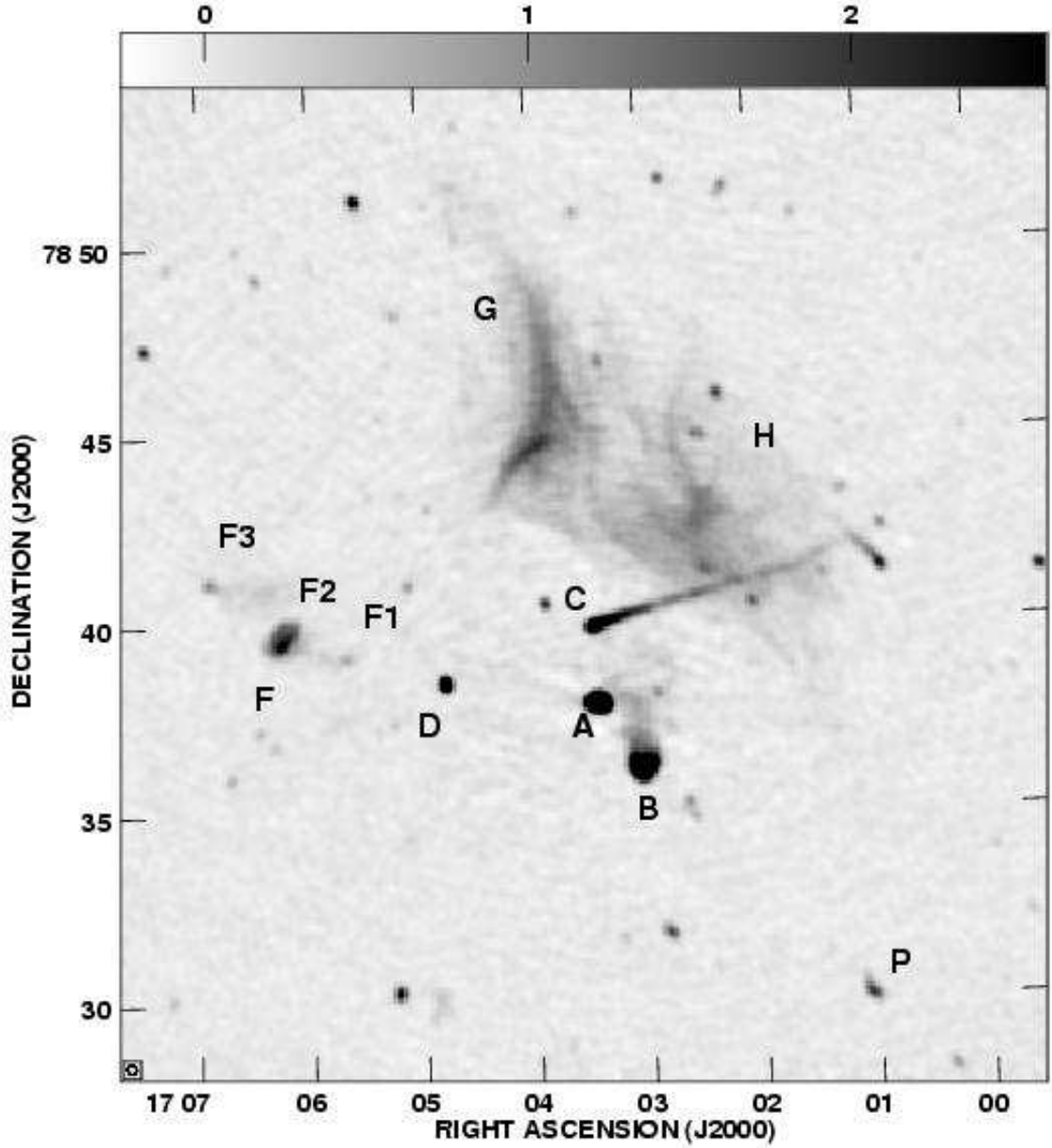}
\caption{VLA C configuration 1369 MHz image of the central region of
Abell 2256. Sources discussed in the text are labeled following the
notation of BF76. The large scale diffuse halo and relic emission is
resolved out due to a lack of short interferometer spacing in the C
configuration data. In addition to the the compact and extended
sources, the relic region is seen to contain a number of synchrotron
filaments. The rms level is 37 $\mu$Jy beam$^{-1}$ and the beam is
$\sim$ 15\arcsec $\times$ 14\arcsec.
\label{fig:filamentsb}}
\end{figure}

The radio relic emission is seen to concentrate into two bright
regions (G and H) which are surrounded by an extended region of
diffuse emission. Our new observations are sensitive to much lower
surface brightness emission than those of \citetalias{rsm} and show
that the relics are even more extensive than previously known. The
entire relic region covers an area of roughly 16\farcm9 $\times$
7\farcm8 ($\sim 1125 \times 520$ kpc) and is bounded by a relatively
steep surface brightness drop on all sides, although the southeast
edge appears particularly sharp as pointed out by \citetalias{rsm}. In
the analysis that follows we will discuss a single relic region which
covers sources G and H as well as the extended diffuse emission which
surrounds them. Higher resolution observations of the relic region by
\citetalias{rsm} and \citet{miller03} show that there are a number of
compact radio sources and two head-tail galaxies in the area covered
by the relic emission. In Figure~\ref{fig:filamentsb} we show the VLA
C configuration 1369 MHz image of the cluster. The diffuse emission is
not visible due to the lack of short baselines, but the relic can be
seen to contain a number of bright synchrotron filaments of width
$\sim$ 30 kpc. We also provide source labels in
Figure~\ref{fig:filamentsb} for all sources discussed in the text.

\begin{figure}
\plotone{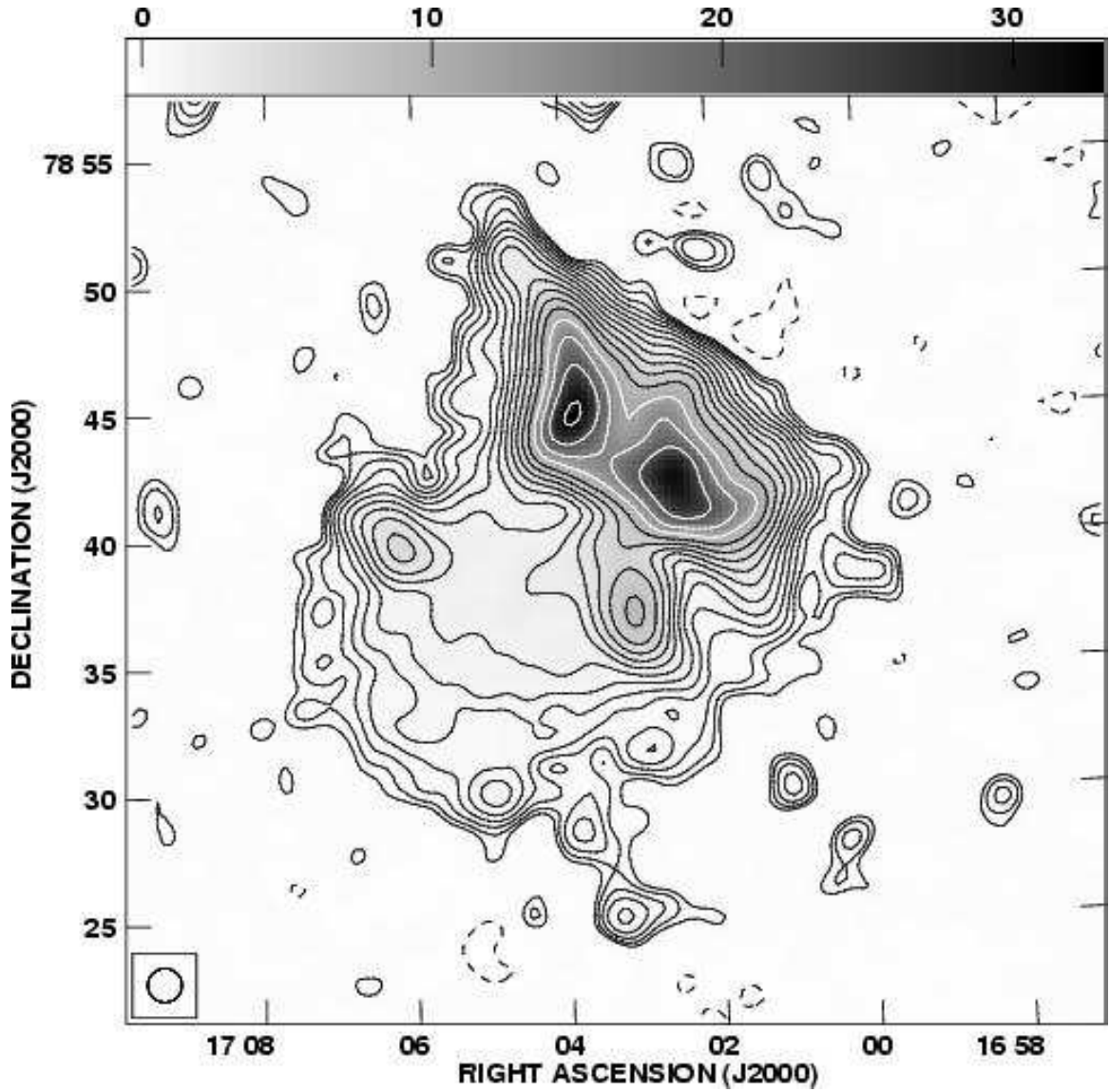}
\caption{VLA 1369 MHz image of Abell 2256 after subtraction of
discrete sources. The resolution is 80\arcsec\ and the noise level is
88 $\mu$Jy/beam, with the contour spacing as in
Figure~\ref{fig:vla_l}. Several of the sources superimposed on the
halo were extended over a large range of spatial scales and thus could
not be completely removed from the image.
\label{fig:diffuse}}
\end{figure}

We have estimated the total flux in the radio relic region using
several different methods in order to account for the influence of the
additional source flux from the discrete sources. Our first method
involved subtracting the total flux of the compact and head-tail
galaxies in the region (using the data from Table 2 of
\citealt{miller03} or Table 3 of \citetalias{rsm}) from the flux of the
region in our D configuration observations. This method may produce an
overestimate of the relic flux as there are several compact sources in
the relic region which do not have flux measurements in
\citet{miller03}. Additionally, the high-resolution data of
\citet{miller03} may resolve out some of the tail of the extended
head-tail source C. This is seen both in a visual comparison of their
Figure 1 with Figure 10 of \citetalias{rsm} as well as by the flux
comparison between the two papers which shows $\sim$13\% more flux in
\citetalias{rsm}. A second estimate of the relic flux was obtained by
adding up the flux of the compact sources seen in the region of the
relic in our C-configuration observations and removing that from the
total flux in the D-configuration observations. A third estimate was
obtained from the multiresolution clean image where we identified the
delta-function clean components associated with the compact and
head-tail sources in the central regions of the cluster and subtracted
those from the final maps in order to produce a radio map containing
mainly the diffuse radio emission. We note that there is significant
filamentary structure within the relics as well as extended structure
around several of the radio galaxies so a straight filtering of data
at short baselines will not properly meet our goal of removing just
the discrete source components. We show this map in
Figure~\ref{fig:diffuse}. Although this method does a good job of
removing all the compact sources and the majority of the more extended
discrete sources, there are still a few regions of extended emission
associated with sources A, B, and F \citepalias[in the notation
of][]{bf76}. This excess residual emission is not important for
determining the relic flux but is discussed below for the
determination of the halo flux.

All three methods produce remarkably similar flux estimates. Including
only the standard noise error estimate we find a total 1369 MHz flux
of 462 $\pm$ 0.8 mJy for the entire relic region using the average of
the three measurements. Due to measurement uncertainties in removing
the discrete sources we estimate the error in the total flux to be
roughly 2\%\footnote{Note that the flux errors presented here and for
the halo emission only refer to measurement errors from the given maps
and do not include an assessment of errors from the map-making
process.}.

\subsection{Halo Emission}
\label{sect:halo_em}

In addition to the large region of radio relic emission, our new
observations provide the first detailed image of the diffuse central
halo emission in Abell 2256. This emission is roughly centered on
source D as originally pointed out by \citetalias{bf76}. The size and
morphology of the halo emission is difficult to determine due to the
presence of the discrete sources and apparent overlap of the halo and
relic emission regions. Nevertheless, the improved sensitivity of our
observations shows that the halo is significantly larger than seen
by \citeauthor{bf76}. We use the roughly circular region of the halo in
Figure~\ref{fig:vla_l} that runs counter-clockwise from the north-east
to south-west and estimate the radius of the halo emission to be
6\farcm1, or 406 kpc. The edges of the halo are much more diffuse than
those of the relic and it is possible that deeper observations will
reveal that the halo emission is even more extended.

We used similar methods to those described above to estimate the total
flux of the halo and the systematic error due to the removal of
contaminating sources. One difference in the multiresolution clean
estimate of the halo flux was that we corrected for the large residual
flux left in Figure~\ref{fig:diffuse} from sources A, B, and F. This
was done by subtracting the total flux in the areas of excess around
each source and then adding back in a flux estimate for the region
excised based on the average halo flux in the region of each source. 

Due to the large flux contributions and extended sizes of the sources
embedded in the halo, the three methods of estimating the halo flux
show larger scatter. Including the standard noise estimate we find a
total 1369 MHz flux of 103.4 $\pm$ 1.1 mJy for the halo. The
uncertainties in this measurement are on the order of 20\%.

\section{Spectral Index}
\label{sect:spec_ind}

We have undertaken wide-field multiresolution clean imaging of an
additional three frequencies from our D-configuration
observations. These images are shown in panels b (1417 MHz), c (1512.5
MHz), and d (1703 MHz) of Figure~\ref{fig:vla_l}. The radio relic
emission shows very little change across the four frequencies with the
only visible difference being the loss of the small low surface
brightness extension to the north-east of the eastern relic edge in
the higher frequency maps. The halo shows significant change across
the frequencies with the majority of the southern portion of the halo
missing in the 1703 MHz observations. Unfortunately, due to radio
frequency interference at 1703 MHz the noise level in our highest
frequency map is roughly 13\% higher than that of the other 3
frequencies. We note also that the $uv$ coverage across the four
frequencies changes resulting in a loss of sensitivity to the largest
scale emission in the 1703 MHz data. It is quite possible that the
missing southern portion of the halo emission at 1703 MHz is the
result of these instrumental limitations. 

\begin{figure}
\plotone{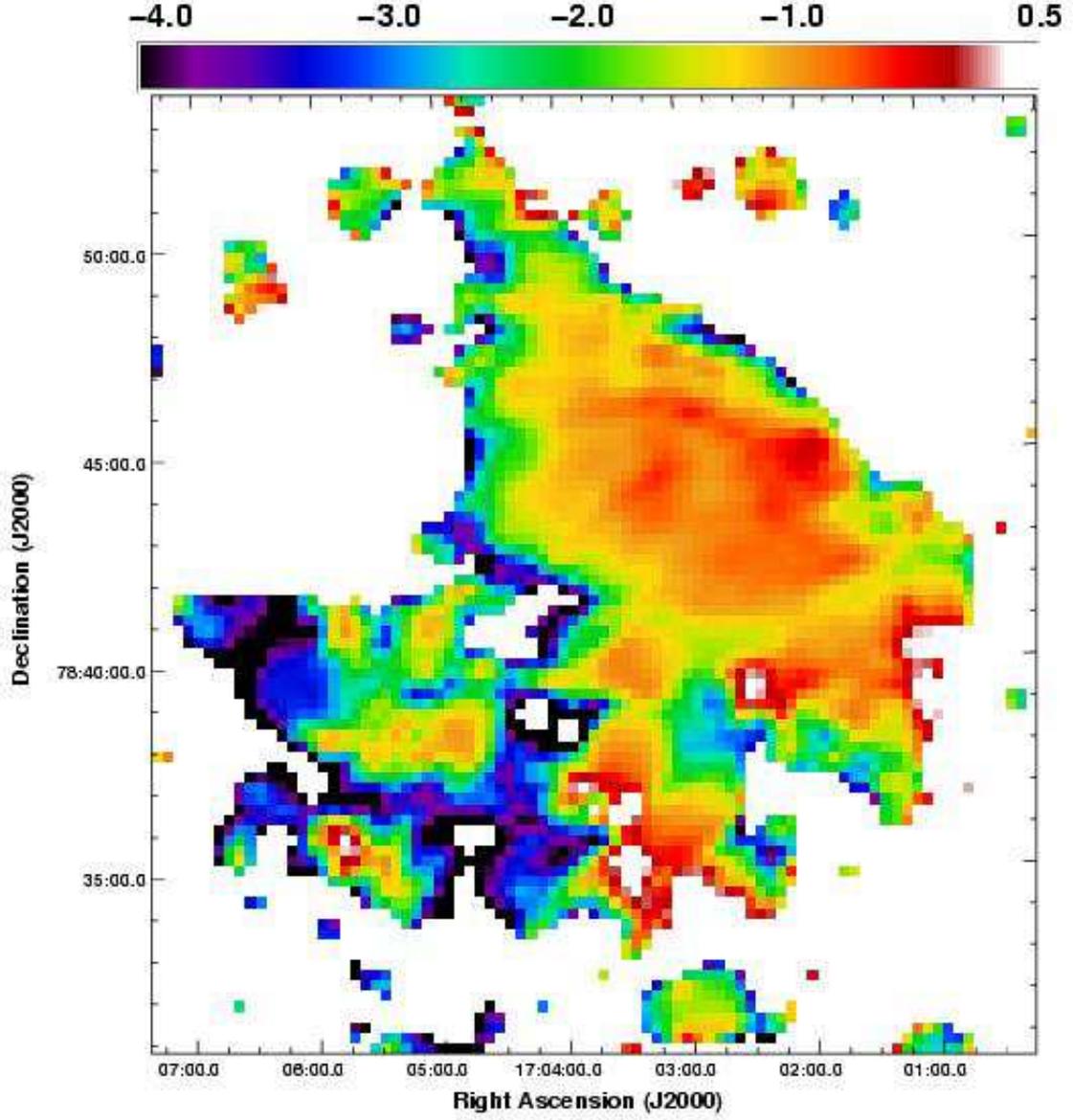}
\caption{Spectral index map between 1369 MHz and 1703 MHz. The color
bar at the top runs from spectral indices of -4.0 to +0.5 where $S_\nu
\propto \nu^\alpha$. The spectral index of the relic region shows a
steepening from the northwest to the southeast. Note that the spectral
steepening across the tail of source C is also seen as a linear
feature in the southwest portion of the map. We note that the spectral
index over the diffuse halo is likely affected by the $uv$ coverage
issues discussed in \S~\ref{sect:spec_ind} and should not be
estimated from this map.
\label{fig:Lspix}}
\end{figure}

To make the spectral index map we smoothed the individual total
intensity maps with a circular 53\arcsec\ Gaussian beam. Due to the
large size of Abell 2256, each map was corrected for the attenuation
of the primary beam of the VLA antennas. After computing the spectral
index map, it was blanked anywhere that the noise on either of the two
input maps fell below the 3$\sigma$ level ($\sigma_{1369}=1.0\times
10^{-4}$ Jy, $\sigma_{1703}=1.1\times 10^{-4}$ Jy). We show the
spectral index map between 1369 MHz and 1703 MHz in
Figure~\ref{fig:Lspix}. Overall the spectral index of the relic region
is fairly uniform with an average spectral index ($S_\nu \propto \nu^\alpha$) of $\alpha=-1.2$. The
extended tail of source C appears as a steeper spectrum region in the
figure. The spectral steepening along the tail of this source is
consistent with the measurements of \citetalias{rsm}. Assuming a uniform
spectral index of $\alpha=-1.2$ for the
relic, we estimate the total 1.4 GHz rest-frame power of the relic
region is $P_{1.4}=3.6\times 10^{24}$ W Hz$^{-1}$.

\begin{figure}
\plotone{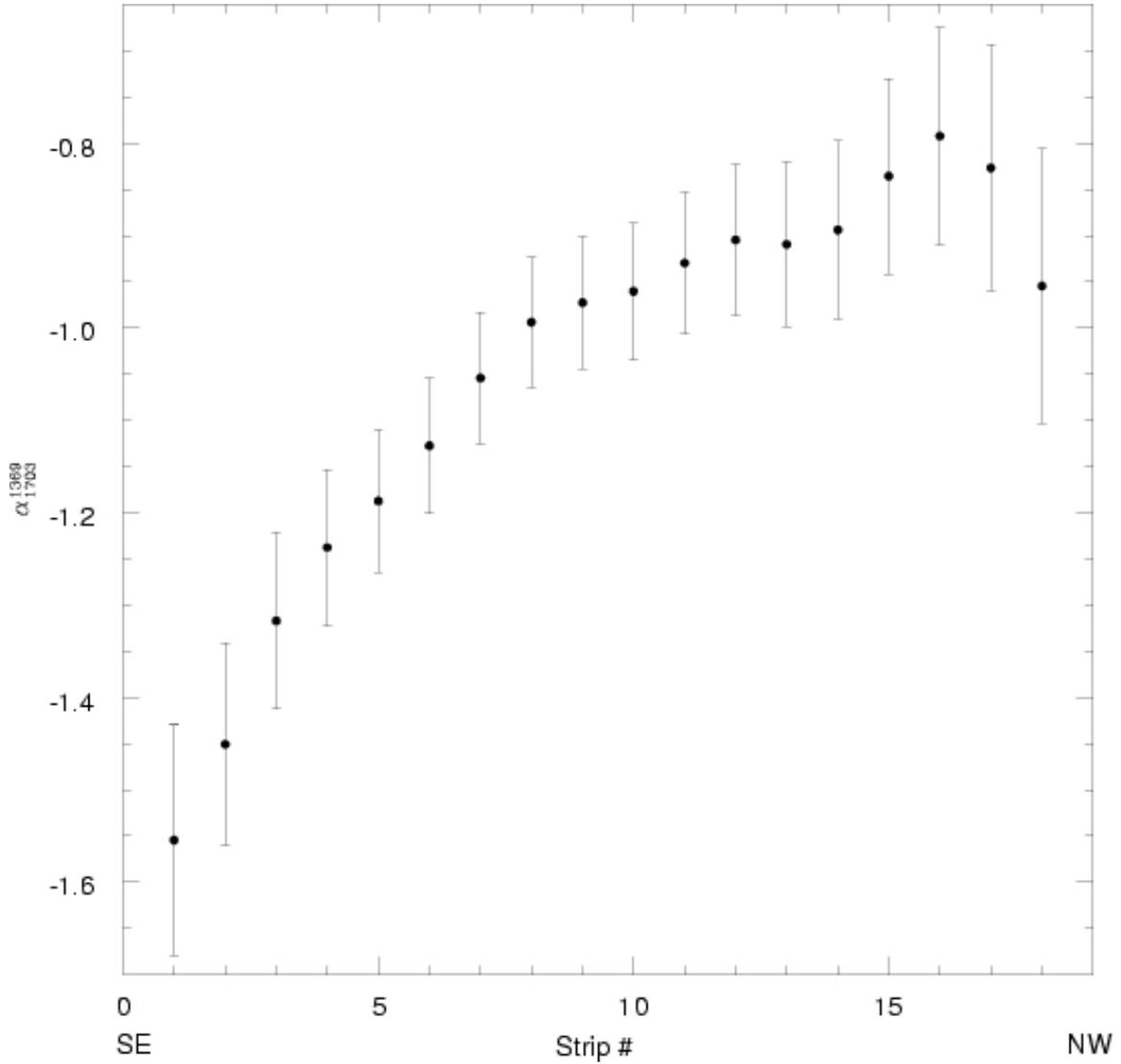}
\caption{Average spectral index in 18 strips of width 6\farcm5
parallel to the long edge of the relic. The average spectral index in
each strip is plotted for strips 1 through 18 where strip 1 is near
the SE relic edge, and strip 18 is near the NW relic edge. Note that
while there are 18 strips plotted with the spectral index they are not
all independent as there are 3 pixels per beam and thus roughly 6
independent resolution elements across the width of the region used to
estimate the spectral index profile. The error bars show the 1$\sigma$
errors on the spectral index.
\label{fig:proj}}
\end{figure}

There is some evidence for a spectral steepening from the NW edge of
the relic toward the SE edge. To look at the trend of spectral index
with location across the relic we have set up 18 strips of width
6\farcm5 parallel to the long edge of the relic (i.e.\ at a position
angle of 53\degr). We have determined the average spectral index in
each of the strips and plot these in Figure~\ref{fig:proj} where strip
1 is located toward the SE edge of the relic and strip 18 is near the
NW edge. The first eight strips (covering roughly 2\farcm5) show a
roughly linear increase in spectral index from $\alpha=-1.5$ to
$\alpha=-1.0$. The remainder of the strips (covering $\sim$ 3\arcmin)
show relatively constant spectral index of $\alpha\sim -0.98$. A
careful examination of Figure 4 of \citet{bf79} shows a similar
spectral trend along the relics between frequencies of 1415 MHz and
610 MHz. We note that the spectral index we measure for the relics
between 1369 MHz and 1703 MHz is significantly steeper than that
reported by \citetalias{rsm} between 1446 MHz and 327 MHz and suggests a
spectral flattening of the relics to lower frequencies. We will
undertake a more detailed spectral analysis of the diffuse emission in
a subsequent paper where we present data over a much larger range of
wavelengths. We defer a detailed discussion of the spectral index
implications to that paper.

The diffuse radio halo is a low surface brightness source with several
large bright sources superimposed. Across the short frequency baseline
presented in this paper it is very difficult to measure the halo flux
with sufficient accuracy to estimate the spectral index. We discuss
above the effects of $uv$ coverage on the diffuse emission, and in
\S~\ref{sect:halo_em} we estimate the uncertainty in the 1369 MHz halo
flux is on the order of 20\%, thus we defer a new measurement of the
halo spectral index to a later paper. Using a longer frequency
baseline between 151 MHz and 610 MHz, \citet{bf79} estimate the halo
spectral index is $\alpha=-1.8$. Using this spectral index together
with our measured 1369 MHz flux, we estimate the total 1.4 GHz
rest-frame power of the halo region is $P_{1.4}=8.2\times 10^{23}$ W
Hz$^{-1}$.

\begin{figure}
\plotone{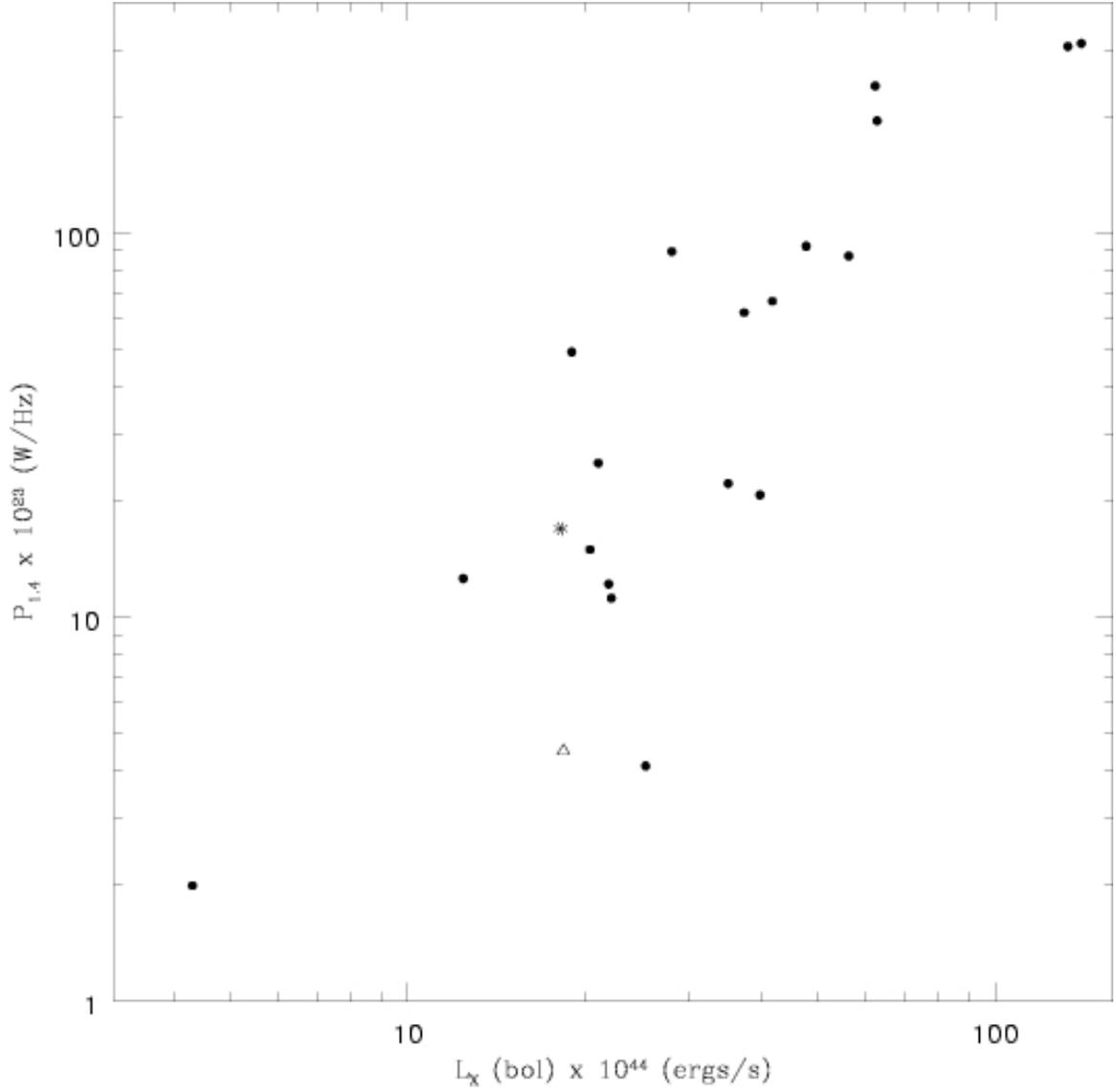}
\caption{Plot of the $L_X-P_{1.4}$ relation for cluster radio
halos. Filled black circles show halos from \citet{govoni01},
\citet{feretti02}, \citet{bacchi03}, and \citet{venturi03}. The star
shows the A2256 halo flux from this work and the open triangle is the
A2256 halo flux reported in \citet{feretti02}. Note that we follow the
convention in the literature for halo studies and adopt $H_0=50$ km
s$^{-1}$ Mpc$^{-1}$, $q_0=0.5$ in this plot.
\label{fig:lx-p}}
\end{figure}

\citet{liang00} showed that the 1.4 GHz rest-frame radio power of
well-confirmed radio halos is correlated with the X-ray luminosity
($L_X$) of the host cluster. We show in Figure~\ref{fig:lx-p} the plot
of the bolometric X-ray luminosity versus 1.4 GHz rest-frame halo
power for known radio halos. Our new measurement of the halo power
for A2256 (star) yields a halo power roughly 4 times higher than the
previous published halo power from \citet{feretti02} (open triangle)
and move A2256 from an outlier position onto the correlation found by
\citet{bacchi03}.

The third steep-spectrum radio source in Abell 2256 is the Z-shaped
source located on the eastern side of the halo \citepalias[source F in
the notation of][]{bf76}. \citet{miller03} find that the eastern
component (F3) is associated with a cluster member galaxy and is
probably not related to the remainder of the source. The central
portion of the source (F2) appears to have a shell-like morphology
\citep{rsm,miller03} and may be a member of the phoenix radio relic
class \citep{taxon} which are shock compressed old radio lobes (so
called radio ghosts) of former radio galaxies. We find a steep
spectral index of $\alpha=-2.5\pm0.2$ for the F2 portion of the source
between 1369 MHz and 1703 MHz.  

\section{Polarization Properties}

The radio observations of Abell 2256 presented in this paper were
taken in full polarization mode to allow us to investigate the
polarization properties of the cluster. For each of the four
frequencies, we have obtained images in all four Stokes parameters. We
have used the Stokes Q and U images to make maps of the total linear
polarization in this system. The polarization percentage map was made
by taking the ratio of the linear polarization map and the total intensity
map and blanking the output map anywhere that either the total
intensity or linear polarization fell below 3$\sigma$. 

\begin{figure}
\plotone{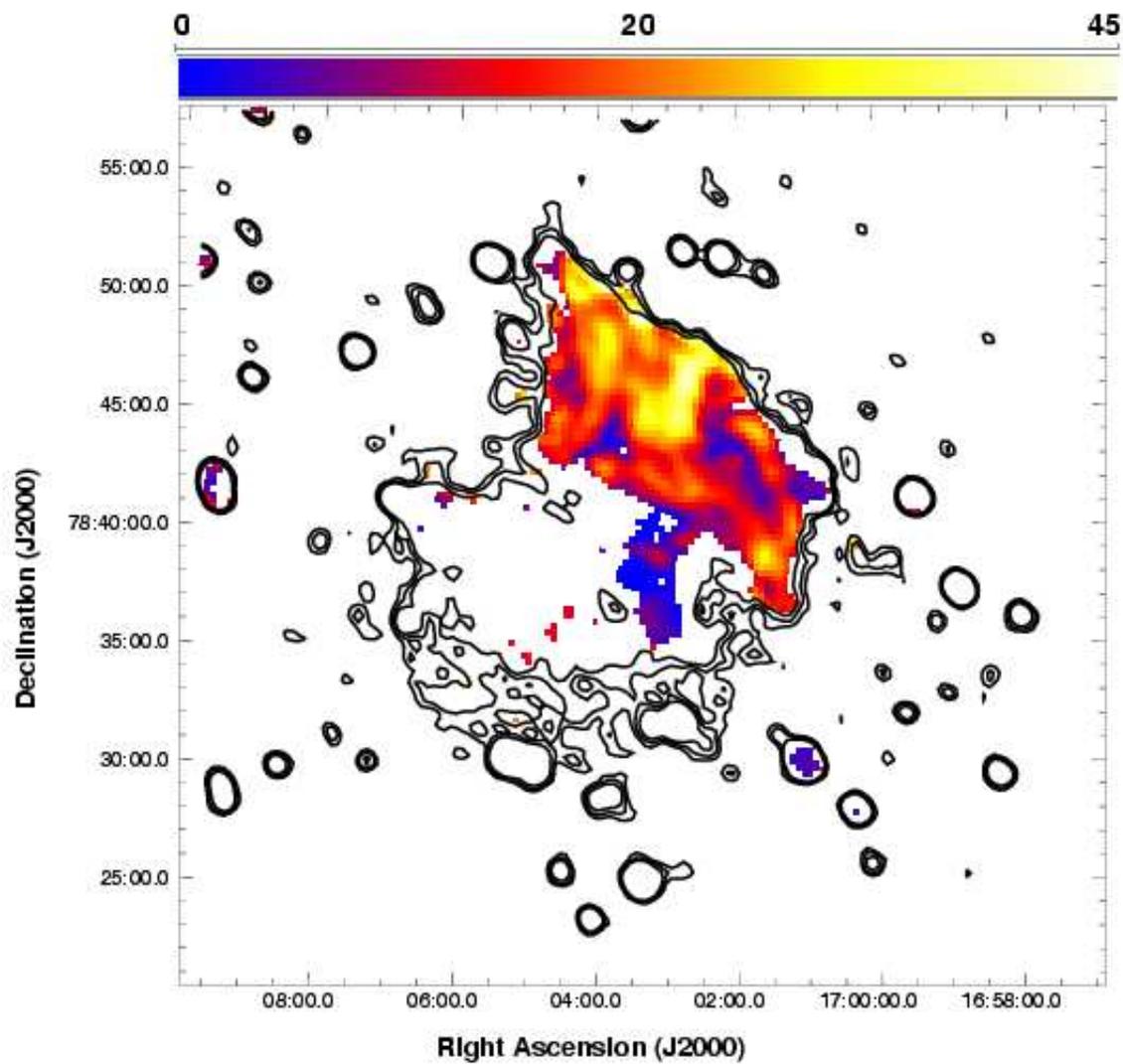}
\caption{Fractional polarization is shown in color with the outer
three contours from the VLA 1369 MHz D configuration data. The color
bar at the top runs linearly from 0\% to 45\% polarization.
\label{fig:Fpol}}
\end{figure}

In Figure~\ref{fig:Fpol} we display the fractional polarization of the
radio emission at 1369 MHz. We have included the outer contours of the
1369 MHz total intensity emission in the plot for comparison with
Figure~\ref{fig:vla_l}. We see that the discrete sources A, B, and
P \citepalias[in the][notation]{rsm} are polarized at the level of
$1-5$\%. This polarization fraction is typical for extragalactic radio
sources observed at this frequency \citep{srs94,m02}. The polarization
map also reveals that the entire radio relic shows high levels of
polarization while there is no detectable polarization associated with
the radio halo emission or the Z-shaped source. 

\begin{figure}
\vskip-0.6truein
\plotone{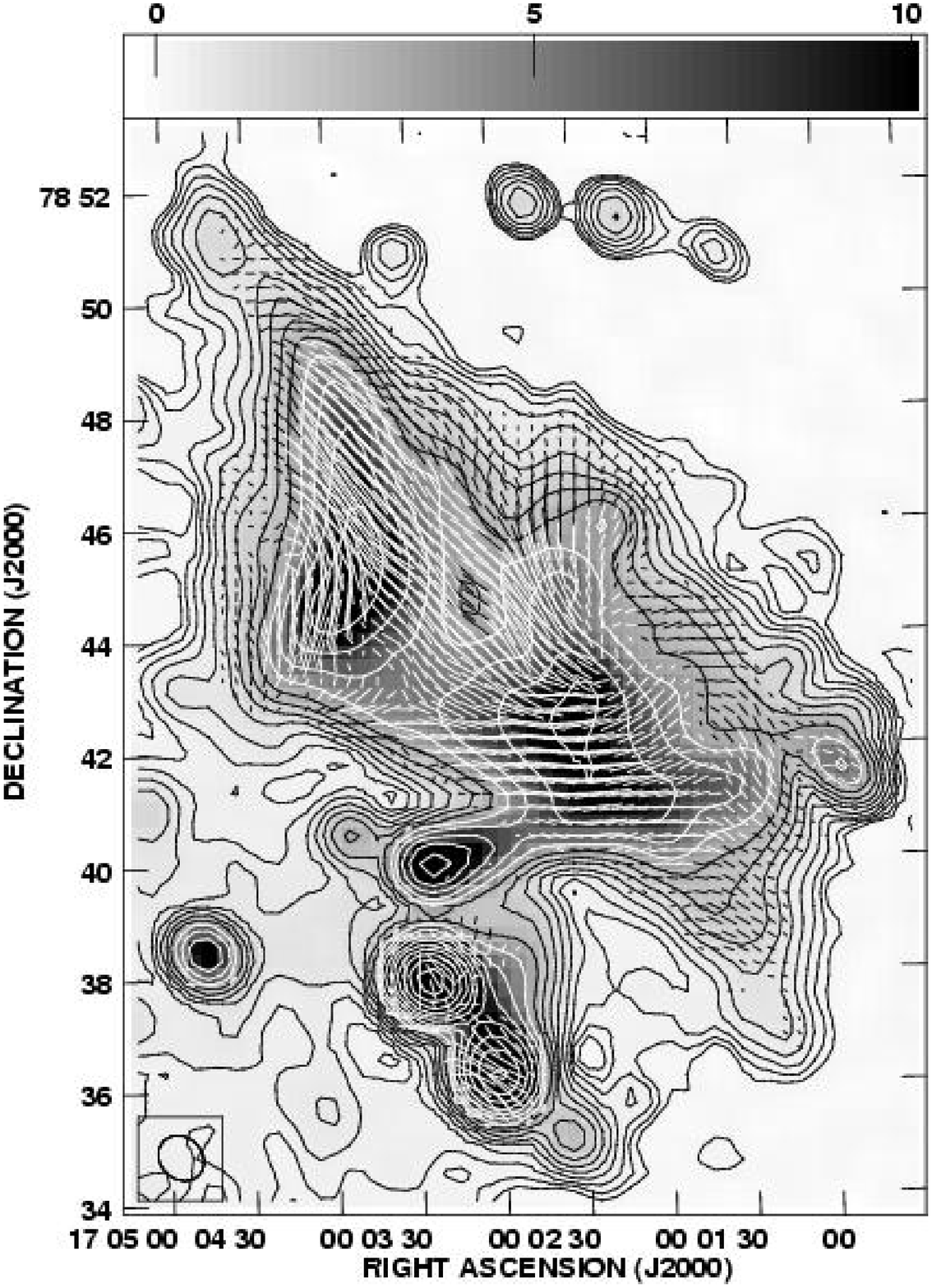}
\vskip0.2truein
\caption{VLA 1369 MHz D configuration contours of the Abell 2256 relic
region with the Faraday corrected magnetic field vectors overlaid. The
polarization shows large regions of uniform direction over more than
half of the length of the relic. Sources A and B (in the
BF76 notation) are also polarized.
\label{fig:bfield_relics}}
\end{figure}

\begin{figure}
\plotone{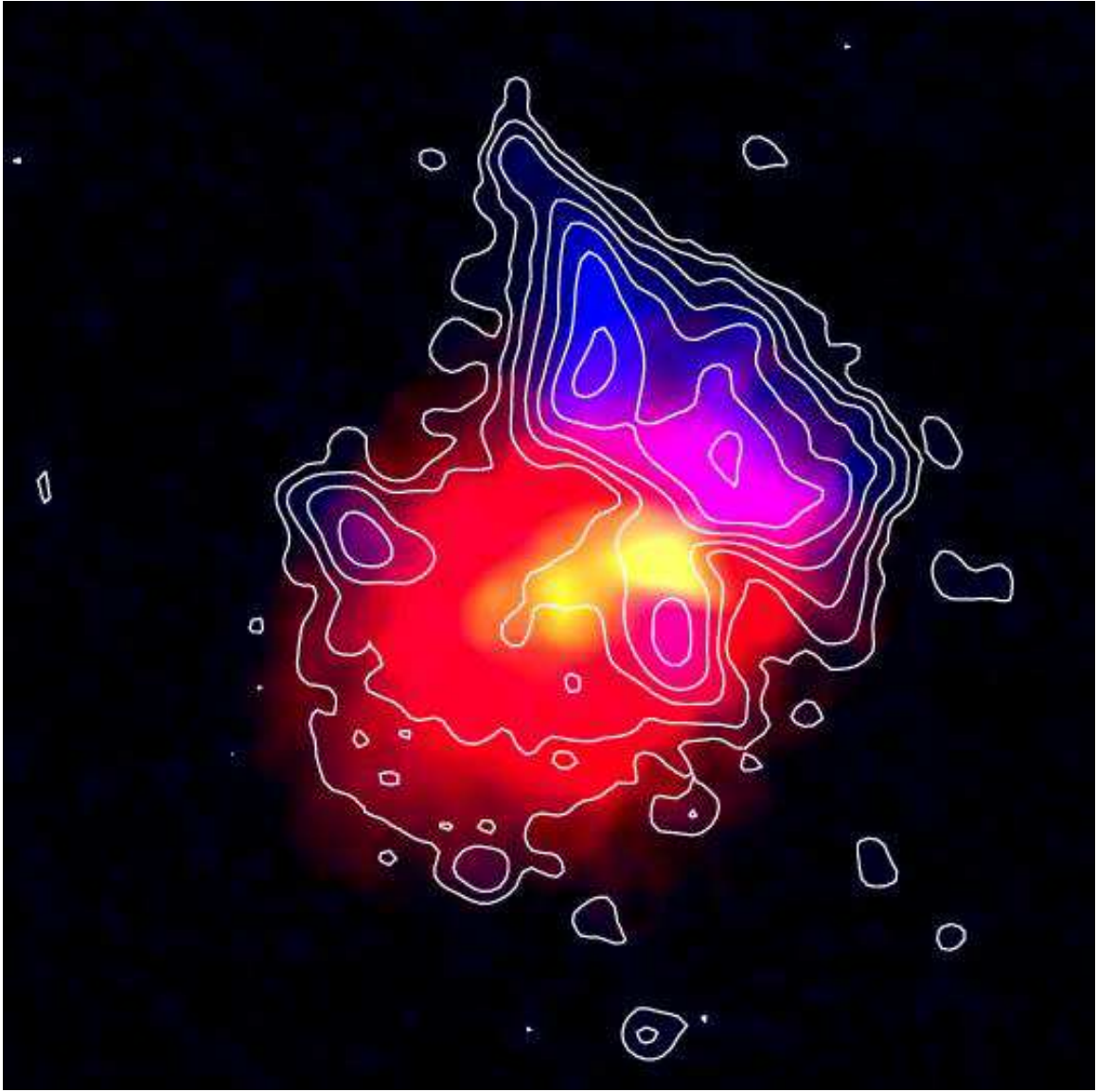}
\caption{Three color image of the radio and X-ray emission in Abell
  2256. Contours and blue emission show the 1369 MHz synchrotron
  emission from Figure~\ref{fig:diffuse}, while the $Chandra$ X-ray
  image is shown in red and green to reveal both the extended diffuse
  thermal emission as well as the three compact X-ray concentrations
  discussed in \citet{sun}. The large X-ray structure near the
  southeast edge of the relic is the cold core of the merging
  subcluster which is thought to be responsible for creating the diffuse radio emission.
\label{fig:chan_over}}
\end{figure}

The linear polarization of the radio relic emission reaches a
fractional polarization of up to 45\% in the interior region with an
average polarization over the entire relic of 20\%. This average
polarization fraction is consistent with the value reported by
\citet{bf79}. There are two extended regions in the relic where the
polarization drops to the $3-5$\% level (blue regions in
Figure~\ref{fig:Fpol}). In Figure~\ref{fig:bfield_relics} we show the
region of the relic emission with the Faraday-corrected magnetic field
vectors plotted on the total intensity greyscale and contour image at
1369 MHz. At the resolution of these images, the magnetic fields show
large scale order over distances of up to 7\arcmin\ (475 kpc). The
magnetic field over a large region in the central part of the relic is
oriented roughly parallel to the long axis of the emission. The
polarization vectors show a change in the field orientation in the
brightest region of the western portion of the relic near the
head-tail galaxy C. This region coincides with the location of the
bright X-ray feature seen around the region P$_2$ in the notation of
\citet{sun}. We show an overlay of the diffuse radio emission (after
subtraction of discrete sources) and the smoothed X-ray emission in
Figure~\ref{fig:chan_over}.

In any relic formation scenario the magnetic fields become aligned
with the shock plane. That means that seen with some inclination, even
an originally isotropic distribution of fields would exhibit a
preferred direction, observable by a global polarization signature. We
find that the averaged polarized emission is still at the 20\% level
over a larger region of the relic. The averaged polarization B-vector
over the sub-region of the relic used to measure the polarization
fraction is aligned on a position angle of 10\degr\ (measured east of
north). The polarization fraction indicates an inclination between the
shock plane and plane of sky of $\sim$ 45\degr\ \citep{ebkk98} with
the average B-vector lying in both planes. Note however, that since
the relic shows large scale magnetic structures with sizes comparable
to the relic, the assumption of an initially isotropic field
distribution within the relic volume needs not be completely
fulfilled.

Our observations do not show any significant polarization for the
diffuse halo emission. Using the 3$\sigma$ noise level in our
polarization images we place a conservative upper limit of 16\% on the
polarization of the halo. Due to the low surface brightness of the
halo, this upper limit is higher than that for the Coma halo
\citep[$<$ 10\%,][]{fg98}. More strict upper limits on halo
polarization have been obtained for Abell 2219 \citep[$<$
6.5\%,][]{bacchi03}, Abell 2163 \citep[$<$ 4\%,][]{feretti01}, and
$1E0657-57$ \citep[$<$ 6.5\%,][]{liang00}. The first clear detection
of polarization in a halo area has recently been presented by
\citet{govoni05} where a number of filaments polarized at levels of
$20\% - 40\%$ were detected. Similarly, the lack of polarized emission
in the region of the Z-shaped source places an upper limit of 2\% on
its linear polarization.

\begin{figure}
\plotone{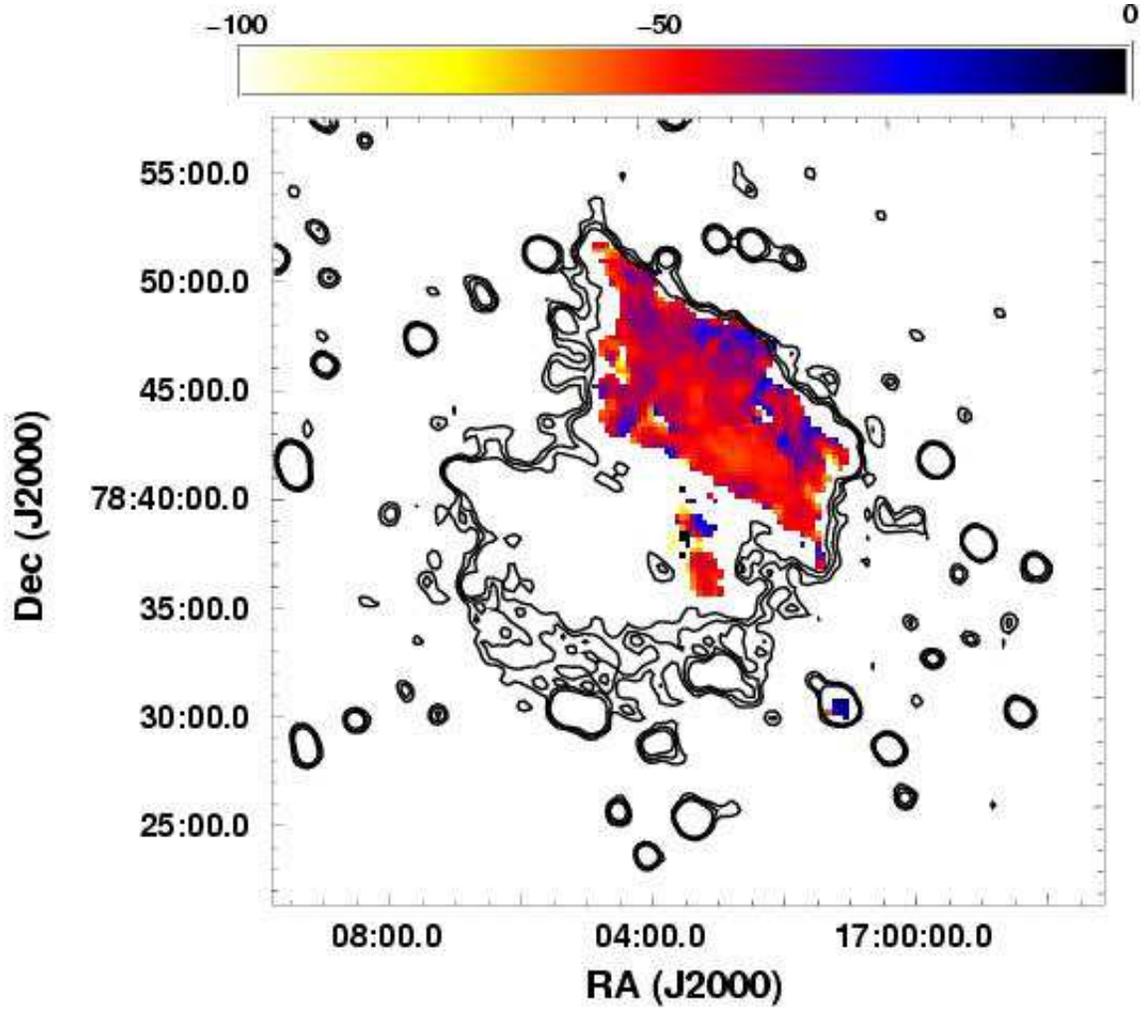}
\caption{Faraday rotation measure map of the radio relics. Contours
show the outer 3 contours from the VLA 1369 MHz D configuration image
to provide a reference. The color scale shows the rotation measure
running linearly from -100 ${\rm rad/m^2}$ to 0 ${\rm rad/m^2}$. The
rotation measure is relatively uniform over the majority of the
relic with a mean rotation measure of -44 rad m$^{-2}$ and a dispersion of 7 rad m$^{-2}$.
\label{fig:RM_map}}
\end{figure}

We have combined the polarization data of all four frequencies to
produce a Faraday rotation measure (RM) map which we show in
Figure~\ref{fig:RM_map}. The map was created by tapering the Stokes Q
and U images at all four frequencies to similar beams then convolving
each image to a 53\arcsec\ $\times$ 45\arcsec\ beam at a position
angle of 45\degr. The rotation measure was calculated using a weighted
fit of the position angle to the wavelength squared. The output map
was blanked anywhere that the position angle errors at any input
frequency were greater than 15\degr. 

The RM map shows a remarkably uniform distribution across the relics
with a mean rotation measure of -44 rad m$^{-2}$ and a dispersion of 7
rad m$^{-2}$. At the position of Abell 2256 ($\ell_{II}=$111\fdg09,
$b_{II}=$31\fdg7) the Galactic RM is expected to be about -4$\pm$ 37
rad m$^{-2}$ based on an average of the 7 sources within 15\degr\ in
the RM catalog of \citet{skb}. The Faraday rotation measure across the
relic is consistent with being Galactic in origin and thus any cluster
component is likely to be very small. This suggests that the relic is
on the near side of the cluster and is not experiencing significant
Faraday rotation from the intracluster medium.

\section{Diffuse radio emission and the geometry of the merger}
\label{sect:geom}

In the following we try to understand the three dimensional geometry of the
merger by comparing the observed properties of the radio halo and relic to
current theoretical concepts of their nature. 

Our spectral index measurements of the radio relic reveal that the
spectrum across the 20 cm band seems to steepen from $\alpha=-0.98$
near the NW edge to $\alpha=-1.5$ near the SE edge. Radio relics in
clusters are believed to be a direct tracer of merger shock
waves. They could be due to Fermi acceleration of relativistic
electrons at a merger shock wave \citep{ebkk98, r99a,
2001ApJ...562..233M} or due to compression and re-ignition of fossil
radio plasma \citep[so called {\it radio ghosts} or {\it ghost
cavities},][]{1999dtrp.conf..275E, eg01, eb01,
2004MNRAS.347..389H}. In the notation of \citet{taxon} the former are
the ``radio gischt'' or ``radio tsunami'', while the latter are the
``radio phoenix''. Given the Mpc size of the relic structure in A2256,
the second explanation is unlikely since the radiative energy losses
during the time it takes to compress a several hundred kpc sized radio
ghost would remove most of electrons responsible for the observable
radio emission. Therefore, it is more likely that we are witnessing
direct shock acceleration in the relic region. In this case, one
expects a trend in the spectral index of the radio emission from
flatter to steeper as one moves from the current shock location to the
trailing edge (in the observers projection) \citep{ebkk98}. The
spectral index profile in Figure~\ref{fig:proj} therefore suggests
that the part of the merger shock wave associated with the relic
probably travels from the SE toward the NW.

Cluster radio halos are also believed to be linked to shock waves from
cluster mergers. There is a significant correlation between the
presence of a radio halo in a cluster and a complex X-ray morphology
or other merger indicators \citep{fg96, buote,
2001A&A...378..408S}. The different theoretical explanations for halos
all seem to favor a developed state of the merger. Magnetic fields are
expected to be stronger during the turbulent phase after a merger has
proceeded past core passage \citep{2002A&A...387..383D,
2005astro.ph..5144S, 2005astro.ph..5517E}. This should lead to
enhanced synchrotron emissivity in the post merger phase in nearly any
radio halo model. In the {\it in situ acceleration model}, a certain
level of turbulence is required to re-accelerate existing relativistic
electron populations to higher energies \citep{1993ApJ...406..399G,
2001MNRAS.320..365B, 2004MNRAS.350.1174B, 2002A&A...386..456G}, also
favoring a post-merger phase. In the case of the {\it hadronic
secondary model} the radio emitting relativistic electrons are
produced by inelastic collisions of relativistic protons with thermal
protons \citep[e.g.][]{1980ApJ...239L..93D, 1982AJ.....87.1266V,
1999APh....12..169B, 2000A&A...362..151D, 2004A&A...413...17P}. A
merger shock wave which has already passed the cluster core is a prime
candidate to have injected the necessary relativistic protons. 

The low level of Faraday rotation measure dispersion ($RM_\mathrm{rms}
\approx 7\,\mathrm{rad/m^2}$) of the relic can also be used to place
constraints on the geometry of the merger. The expected rotation
measure dispersion \citep[see Eq. 40 of][]{2003A&A...401..835E}
increases with the line of sight length $L$
\begin{equation}
\label{eq:RMformula}
\frac{d\langle RM^2 \rangle}{dL}    = \frac{1}{2}\, a_0^2\, n_{e}^2\,
\lambda_B \, \langle B^2  \rangle\,,
\end{equation}
where, $a_0 = {e^3}/({2\pi \,m_e^2\,c^4})$ is the usual Faraday
rotation constant, $n_{e}$ the electron density in the cluster core,
$\lambda_B$ the magnetic autocorrelation length, and $B^2$ the
magnetic field strength. Inserting typical values for the A2256
cluster core electron density and radius from
\citet{1999ApJ...517..627M}, $n_e = 3.5 \times 10^{-3}\,
h_{70}^{1/2}\,\mathrm{cm}^{-3}$, $L = 350\,h_{70}^{-1}$~kpc (a core
radius), with the size of the $RM$ fluctuations equal to the
synchrotron filament width visible in Figure~\ref{fig:filamentsb})
$\lambda_B = 30\,h_{70}^{-1}$~kpc, and a very conservative field
strength of $B_\mathrm{rms} = 1\,\mu\mathrm{G}$ (as discussed in
\S~\ref{sect:ic_field}) yields
\begin{displaymath}
\label{eq:RMformula2}
RM_\mathrm{rms} \approx 205\frac{\mathrm{rad}}{\mathrm{m^2}}\, 
\left(\frac{n_{e}}{3.5\times 10^{-3} \, \mathrm{cm^{-3}}} \right) \,
\left(\frac{\lambda_B}{30\, \mathrm{kpc}} \right)^\frac{1}{2} \,
\end{displaymath}
\begin{equation}
\times \left(\frac{L}{350\, \mathrm{kpc}} \right)^\frac{1}{2} \,
\left(\frac{B_\mathrm{rms}}{1\,\mu\mathrm{G}} \right) \,.
\end{equation}
This estimate is much larger than what is observed, therefore the
rotation measure dispersion favors a merger geometry in which the line
of sight $L$ through the high density ICM is significantly reduced. We
note that the current data do not allow us to measure the $RM$
fluctuation size on smaller scales due to the size of our beam. It
seems unlikely, however that there are large RM fluctuations on very
small scales as that would lead to depolarization within each beam due
to the averaging of different polarization angles across the beam. A
comparison of the polarization fraction of the relics at 1703 MHz with
that at 1369 MHz reveals a depolarization ratio 
($DP=\%P_{1703}/\%P_{1369}$) of 1.08 across the relics which supports
the argument against large RM variations on small scales. It therefore
seems unlikely that we have significantly underestimated the
$RM_\mathrm{rms}$ due to the limited resolution of our data. Using the
best-fit model of \citet{1999ApJ...517..627M} to determine the radial
profile of the electron density, we find that even at the largest
projected radial distance from the cluster core, the electron density
is at most an order of magnitude lower than the central value, and
thus the predicted $RM_\mathrm{rms}$ through the cluster would still
be significantly larger than the observed value. The rotation measure
dispersion, therefore, argues strongly in favor of the relic being
located on the front side of the cluster.

The dynamical state of Abell 2256 is complex and is thought to consist
of at least three merging systems based on optical velocity
dispersions \citep{blc02,miller03} and shows evidence of three X-ray
substructures \citep{sun}. The main optical concentration appears
roughly centered on the primary X-ray concentration, while the low
velocity subcluster is thought to be associated with the cold X-ray
concentration located along the southeast edge of the radio relic. The
third (high velocity) optical system may represent a small group
located to north of the cluster core and has not been detected in
X-rays. The X-ray ``shoulder'' detected by \citeauthor{sun} to the
east of the primary concentration is roughly centered on the halo and
may represent an older merging component or internal structure in the
central region of the primary cluster. The complex dynamical state of
this system makes a unique determination of the merger history very
difficult. We consider the small high velocity northern group to be
unlikely to be related to the origin of the diffuse emission as halos
and relics are nearly always associated with large mergers of systems
with nearly equal mass ratios. 

The X-ray emission associated with the subcluster shows a sharp edge
toward the south which is consistent with a cold front
\citep{sun}. This suggests that at least some of the motion of this
system is toward the south, although \citeauthor{sun} point out that
the flat shape of the edge indicates that we are only seeing part of
the edge due to projection effects. The merger geometry in Abell 2256
may be similar to Abell 754 rotated $\sim$ 180\degr\ in the plane of
the sky. In both Abell 2256 and Abell 754, the cold X-ray core of the
merging component is bounded on one side by a steep surface brightness
gradient. In Abell 754 the X-ray substructure does not appear to be
associated with any large galaxy concentration. \citet{mm03} suggest
that this cold core in A754 may have completely decoupled from its
previous dark matter host and may be ``sloshing'' independently in the
system. It is possible that a similar decoupling has occurred in Abell
2256 (M.\ Markevitch 2005, private communication) where the galaxy
concentration thought to be associated with the merging subcluster is
located to the north of the cold core. The morphology of the system
suggests that the subcluster core may be traveling on a complex spiral
trajectory about the primary cluster. Such a trajectory could result
in the radio halo emission as well as relic location on the front-side
of the cluster with the shock propagation direction moving toward the
northwest as suggested by the spectral index variation across the
relic.  The Z-shaped source may be an additional relic which has been
revived by this merger. While the above merger scenario appears
plausible to explain the observational parameters of A2256, we note
that other scenarios could also be possible and detailed hydrodynamic
simulations would be required to determine the validity of any merger
scenario.

\begin{figure}[t]
\begin{center}
\resizebox{\hsize}{!}{\includegraphics{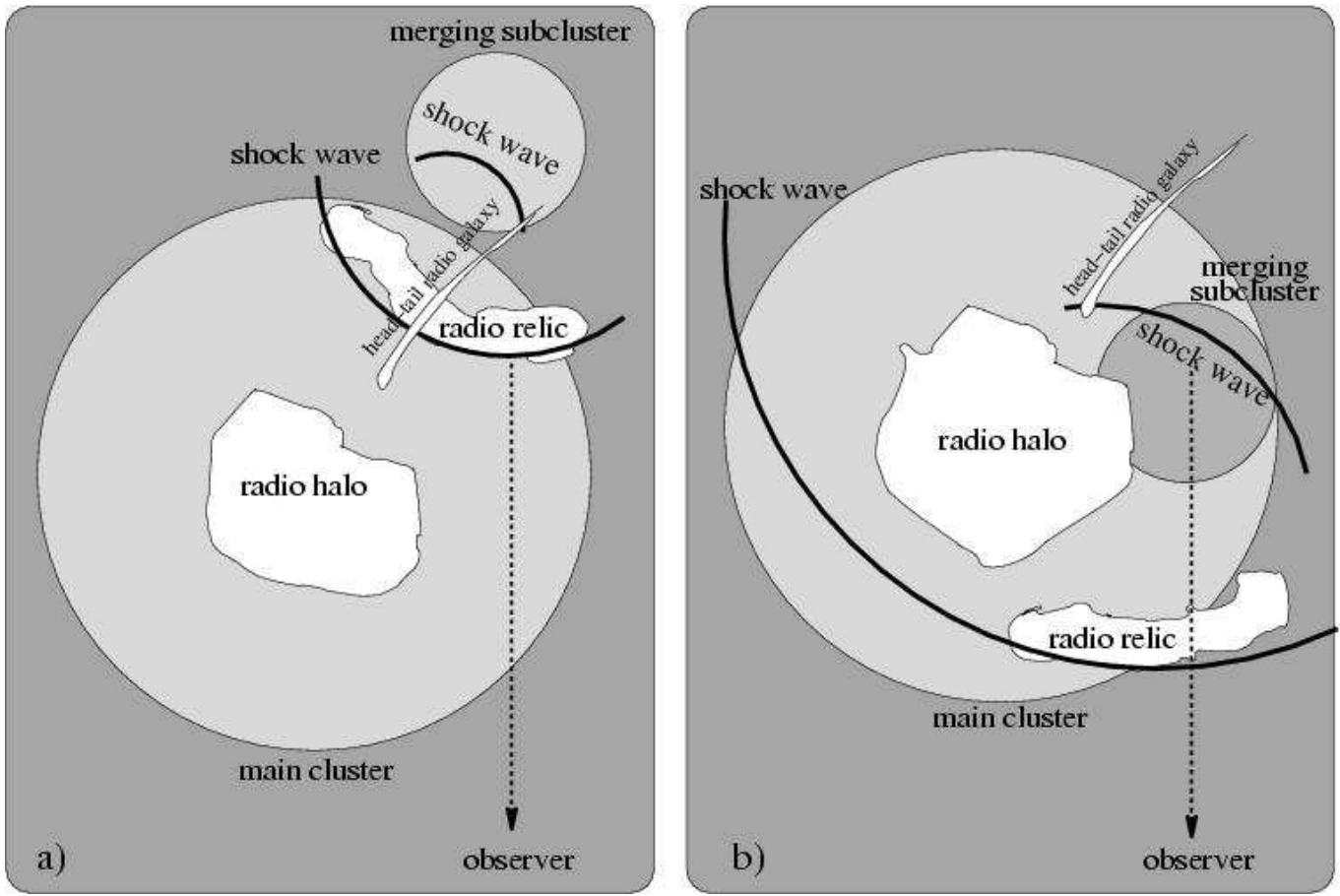}}
\end{center}
\caption[]{Possible geometries of the merger. Scenario a)
  is an early stage of the merger, where the shock waves have not had
  time to pass over the cluster cores, whereas scenario b) is in a
  more developed stage.}
\label{fig:sketchGeo} 
\end{figure}

The two simplest merger scenarios for this system (shown in
Figure~\ref{fig:sketchGeo}) are: a) a current merger in a very early
stage creating the radio relic with the radio halo being the remnant
from an older merger event, or b) an advanced merger between the
primary cluster and western subcluster where the merger shock has
already passed the core of the primary to create both the diffuse halo
in the volume over which the shock wave has swept and relic emission
at the present location of the shock wave. We favor the scenario where
the current major merger is responsible for both the diffuse halo and
relic emission but we note that without detailed hydrodynamic
simulations the current data do not allow us to exclude either model.

\section{Intracluster Magnetic Field}\
\label{sect:ic_field}

The original goal of the polarimetry observations of Abell 2256 was to
use the highly polarized radio relic as a screen against which to
measure the intracluster magnetic field strength and scale. The
observations revealed a small rotation measure of -44 rad m$^{-2}$
which is consistent with the Galactic estimates. At the longitude of
A2256 ($\ell=110^\circ$) Faraday rotation measure maps show a large
region of negative rotation measures running from latitudes of
-40\degr\ up to at least +10\degr\ \citep{skb,jh04}, thus it seems
likely that the observed RM of A2256 at a latitude of +32\degr\ is
related to this feature, and not the intracluster field. Further
support for the Galactic origin of the observed RM comes from the very
low RM dispersion across the relic (see
\S~\ref{sect:geom}). Many more RM measurements in the region of
Abell 2256 would be required to better constrain the Galactic RM
contribution and determine the cluster contribution. We therefore do
not attempt to measure an intracluster magnetic field strength using
the Faraday rotation measure.

The magnetic field strength in clusters of galaxies can also be
estimated through minimum energy arguments applied to the diffuse
radio emission. We apply two different versions of such arguments.
First, a conservative field estimate within the halo results from the
{\it classical minimum energy condition}, which requires that the
total energy in relativistic electrons and protons (assumed to be
proportional to the electrons with a fixed ratio $k_p$) plus the
magnetic energy has a minimum, with the constraint that the observed
radio emission can be generated via the synchrotron process. This
estimate can be applied in any scenario and can be regarded as a lower
limit to the field strength (for conservative choices of $k_p$) unless
one is willing to accept that there is much more energy in
relativistic particles than in magnetic fields.  Second, if radio
halos are generated via the hadronic interactions of cosmic ray
protons with the thermal gas, the proton-to-electron ratio $k_p$ is
not free any more, but is fully determined by the physical processes
such as the hadronic production, and the synchrotron and inverse
Compton cooling of the produced electrons.  This permits us to
calculate a field estimate based on a {\it hadronic minimum energy
condition}. The resulting field strength is only meaningful if the
radio electrons are indeed of hadronic origin. The field estimate
resulting from the hadronic condition will always be above that of
the classical condition with a low adopted value for $k_p$.

We apply both the classical and the hadronic minimum energy criteria
in the {\it nutshell} formulation of \citet{2004MNRAS.352...76P} to
the radio halo of A2256. We adopt the following parameters: a
conservative $k_p =1$, a lower electron energy cut-off of
$E_{e,\mathrm{min}} = 0.1 \, \mathrm{GeV}$ (for the classical case), a
cluster core radius of $350 \,h_{70}^{-1}\,$kpc, a central electron
density of $3.5\times 10^{-3}\,h_{70}^{1/2}\, \mathrm{cm^{-3}}$
\citep{1999ApJ...517..627M}, and a lower proton kinetic energy cutoff
of $0.1 \, \mathrm{GeV}$ (for the hadronic case). We note that the
true spectrum of the halo is expected to display spectral curvature
but this cannot be measured with current data, thus the results are
given for three assumed spectral indices of the radio halo, namely
$\alpha = -1.25$, $\alpha = -1.5$, and $\alpha = -1.7$. Confidence
intervals are computed as described in \citet{2004MNRAS.352...76P} and
mark the region within which the total energy of relativistic
particles and fields stay within one $e$-folding from its minimum.

The classical minimum energy condition yields $B =
1.5_{-0.6}^{+0.9}\,\mu$G, $B = 2.4_{-0.8}^{+1.3}\,\mu$G, and $B =
3.2_{-1.1}^{+1.7}\,\mu$G in the cases $\alpha = -1.25$, $\alpha =
-1.5$, and $\alpha = -1.7$, respectively. The hadronic minimum energy
condition yields $B = 3.3_{-1.2}^{+2.0}\,\mu$G, $B = 5.5_{-2.0}^{+3.1}
\,\mu$G, and $B = 8.9_{-3.1}^{+4.8}\,\mu$G in the cases $\alpha =
-1.25$, $\alpha = -1.5$, and $\alpha = -1.7$, respectively. Due to the
expected curvature of the halo spectrum, the lower magnetic field
values are more likely representative for A2256. Both sets of results
are in good agreement with Faraday rotation measurements of the
magnetic field strength in galaxy clusters \citep{ckb02, ct02}, and
support our above scenario of a radio relic located in the cluster
foreground on the basis of the weak Faraday dispersion for a $\mu$G
level field.

The classical minimum energy fields we find are a bit higher than
typical {\it equipartition} field values of $0.3\,(1+k_p)^{2/7}\,\mu$G
for radio halos given in the literature
\citep[e.g.][]{1993ApJ...406..399G, 1999JKAS...32...75K,
2003A&A...397...53T}. For A2256 \citet{1999JKAS...32...75K} report $
0.3\,(1+k_p)^{2/7}\,\mu$G, which is definitively lower than our
estimate of at least $1\,\mu$G. This difference is due to an important
subtlety. In traditional equipartition calculations the lower cut-off
of the electron spectrum is assumed to correspond to a frequency
cutoff of 100 MHz, which corresponds for $\mu$G field strength to
electron energies of 1 GeV, and to weaker field strength for even
higher electron energies. However, there is no reason why the electron
spectrum should cut off at such energies, or why it should know about
our radio observational window (which closes shortly below 100
MHz). Therefore, we adopted an electron spectral cutoff at 0.1 GeV,
which is motivated by the onset of effective Coulomb cooling below
this energy as pointed out by \citet{1999ApJ...520..529S}. Having
adopted a lower cutoff implies the presence of a larger non-thermal energy
reservoir, which results in a somewhat higher field strength.\\[3em]

\section{Summary and Discussion}

We have presented new radio images of the diffuse synchrotron emission
in A2256. In addition to the well-known radio relic region in this
system, our new observations permit us to obtain the first detailed
maps of the Mpc scale diffuse central halo emission.

Our polarization observations of this system show that the relic
region has a high linear polarization fraction of up to 45\% with an
average polarization across the relic of 20\%. The relic region
contains extended polarization structures with nearly uniform field
direction over more than half the length of the relic. We do not
detect polarization associated with the radio halo or the steep
spectrum Z-shaped source, and place upper limits on their polarization
fraction of 16\% and 2\%, respectively. Our current data are limited
to low resolution in order to be sensitive to the low surface
brightness emission of the halo. If there are magnetic structures in
the halo on scales smaller than our resolution ($\sim 50$ kpc) the
averaging of these structures in our beam could result in the low
observed fractional polarization.

The Faraday rotation measure across the radio relic is remarkably
uniform with a mean rotation measure of -44 rad m$^{-2}$ which is
consistent with our best available Galactic estimate from sources from
the catalog of \citet{skb} within 15\degr\ of the cluster. The low
mean $RM$ and very low dispersion (7 rad m$^{-2}$) across the large
region covered by the relic ($\sim$ 1125 $\times$ 520 kpc) both
support a geometry with the relic located on the foreground side of
the cluster with very little RM contribution from the intracluster
medium. These data, as well as those of \citet{perseus05} on Perseus,
represent the first opportunities to use radio polarimetry to
determine the line-of-sight position of a merger shock wave in a
galaxy cluster.

The radio relic region was found to have an average spectral index of
$\alpha=-1.2$ between 1369 MHz and 1703 MHz with a radial steepening
toward the cluster core. In the relic formation models, the diffuse
emission is connected to shock waves generated by merger events. Given
the large scale of the relic, it seems most likely that the emission
is due to direct shock acceleration where the most recent acceleration
(marking the the current shock location) is expected to be associated
with the flattest spectrum emission. This suggests that (projected on
the sky) the shock has moved from the southeast toward the northwest
across the relic. It seems likely that the two diffuse non-thermal
emission regions, the halo and the relic, are both related to the
current violent hydrodynamical event. This would require the merger to
be in an advanced stage where an associated merger shock has already
progressed past core passage. We note, however, that detailed
hydrodynamical simulations would be required to confirm this.

\acknowledgments We thank the anonymous referee for helpful
suggestions, Christoph Pfrommer for providing us with his minimum
energy code, and Ming Sun for providing us with the $Chandra$ image of
Abell 2256. We thank Bruno Deiss, Uli Klein, Phil Kronberg, Maxim
Markevitch, Neal Miller, Ming Sun, and Michael Thierbach for helpful
discussions. TEC acknowledges support for this work from the National
Aeronautics and Space Administration, primarily through {\it Chandra}
awards GO4-5149x and GO4-5133x issued by the $Chandra$ X-ray
Observatory, which is operated by the Smithsonian Astrophysical
Observatory for and on behalf of NASA under contract NAS8-39073. Basic
research in radio astronomy at NRL is supported by the office of Naval
Research.

\clearpage

\clearpage

\clearpage 

\clearpage

\clearpage

\clearpage 

\clearpage

\clearpage 

\clearpage

\clearpage

\clearpage

\clearpage

\begin{deluxetable}{lclll}
\tabletypesize{\scriptsize}
\tablecaption{VLA Observations. \label{tbl:data}}
\tablewidth{0pt}
\tablehead{
\colhead{Date} & \colhead{Array}   & \colhead{Frequency}   &
\colhead{Bandwidth} &
\colhead{Time}\\
\colhead{} & \colhead{} & \colhead{MHz} & \colhead{MHz} & \colhead{h}
}
\startdata
1999 April 28 & D & 1369/1417 & 25/25 & 6, 6\\
1999 April 29 & D & 1512.5/1703 & 25/12.5 & 5, 3.5\\
2000 May 29 & C & 1369/1417 & 25/25 & 3, 3\\
2000 May 29 & C & 1512.5/1703 & 25/12.5 & 3.5, 3.5\\
2000 June 18 & C & 1369/1417 & 25/25 & 3, 3\\
2000 June 18 & C & 1512.5/1703 & 25/12.5 & 3.5, 3.5\\

 \enddata

\end{deluxetable}

\end{document}